\documentclass[a4paper,11pt,twoside]{amsart}
\usepackage[top=5.cm, bottom=5.0cm, left=3cm, right=3cm]{geometry}
\usepackage[UKenglish]{babel}
\usepackage{times}
\usepackage{hyperref}
\usepackage[foot]{amsaddr}
\usepackage{amsmath}
\usepackage{amsthm}	
\usepackage{amsfonts}
\usepackage{amssymb}
\usepackage{array}
\usepackage{paralist}
\usepackage{algorithmic,algorithm}
\usepackage{float}
\usepackage{graphicx}
\usepackage{adjustbox}
\usepackage{enumerate}
\usepackage{breakcites}
\usepackage[utf8]{inputenc}
\usepackage{dsfont}
\usepackage{ulem}

\theoremstyle{plain}
\newtheorem{theorem}{Theorem}[section]

\newtheorem{proposition}[theorem]{Proposition}

\newtheorem{definition}[theorem]{Definition}

\theoremstyle{definition}

\numberwithin{equation}{section}
\numberwithin{figure}{section}

\usepackage[colorinlistoftodos,bordercolor=orange,backgroundcolor=orange!20,linecolor=orange,textsize=scriptsize]{todonotes}
 
\DeclareMathOperator*{\argmin}{arg\,min}

\title{Solving path dependent PDEs with LSTM networks and path signatures}
\author[M. Sabate-Vidales]{Marc Sabate-Vidales$^1$}
\email{M.Sabate-Vidales@sms.ed.ac.uk}
\author[D. \v{S}i\v{s}ka]{David \v{S}i\v{s}ka$^{1,2}$}
\address{$^1$\href{https://www.maths.ed.ac.uk}{School of Mathematics, University of Edinburgh}}
\address{$^2$\href{https://vega.xyz}{Vega Protocol}}
\email{D.Siska@ed.ac.uk}
\author[L. Szpruch]{Lukasz Szpruch$^{1,3}$}
\address{$^3$\href{https://www.turing.ac.uk}{Alan Turing Institute}}
\email{L.Szpruch@ed.ac.uk}

\date{\today}
\keywords{Monte Carlo method, Deep neural network, Control variates, Stochastic differential equations}

\begin{document}

\setcounter{tocdepth}{1}

\begin{abstract}
Using a combination of recurrent neural networks and signature methods from the rough paths theory we design efficient algorithms for solving parametric families of path dependent partial differential equations (PPDEs) that arise in pricing and hedging of path-dependent derivatives or from use of non-Markovian model,  such as rough volatility models~\cite{jacquier2019deep}.
The solutions of PPDEs are functions of time, a continuous path (the asset price history) and model parameters. 
As the domain of the solution is infinite dimensional many recently developed deep learning techniques for solving PDEs do not apply.  
Similarly as in \cite{vidales2018unbiased}, we identify the objective function used to learn the PPDE by using martingale representation theorem. 
As a result we can de-bias and provide confidence intervals for then neural network-based algorithm. 
We validate our algorithm using classical models for pricing lookback and auto-callable options and report errors for approximating both prices and hedging strategies. 
\end{abstract}

%

\maketitle



\section{Introduction}

Deep neural networks trained with stochastic gradient descent algorithms are extremely successful in number of applications such as computer vision, natural language processing, generative models or reinforcement learning \cite{goodfellow2016deep}. 
As these methods work extremely well in seemingly high-dimensional settings, it is natural to investigate their performance in solving high-dimensional PDEs. 
Starting with the pioneering work~\cite{han2017solving, E_2017}, where probabilistic representation has been used to learn PDEs using neural networks, recent years have brought an influx of interest in deep PDE solvers. 
As a result there is now a number of efficient algorithms for solving linear and non-linear PDEs, employing probabilistic representations, that work in high dimensions~\cite{beck2020overcoming, Berner_2020, gonon2020uniform, grohs2018proof, beck2019deep}.

The methods in~\cite{han2017solving, E_2017} approximate a solution to a single PDE at a single point in space. 
The first paper to extend the above techniques to families of parametric PDEs with approximations across the whole domain was~\cite{vidales2018unbiased}.  
In this paper we extend this to parametric families of path-dependent PDEs (PPDEs). 
Let $B \subseteq \mathbb R^p, p\geq 1$ be a parameter space.
Consider $F=F(t,\omega;\beta)$ satisfying
\begin{equation}\label{eq ppde intro}
\begin{split}
& \bigg[\partial_t F + b \nabla_\omega F + \frac{1}{2}\text{tr}\left[\nabla_{\omega}^2 F \sigma^*\sigma\right] - rF\bigg](t,\omega;\beta)  = 0\,, \\
& F(T,\omega;\beta)  = g(\omega;\beta)\,,\,\,\, t \in [0,T]\,,\,\, \omega \in C([0,T];\mathbb R^d)\,,\,\,\beta \in B\,.
\end{split}
\end{equation}
Here $t\in [0,T]$, $\omega \in C([0,T];\mathbb R^d)$ and $\beta\in B$ and $b,\sigma, r$ and $g$ are functions of $(t,\omega;\beta)$ which specify the problem.
Notice that we are dealing with an equation in an infinite dimensional space.
The derivatives $\nabla_\omega$ and $\nabla_\omega^2$ are derivatives on the path space $C([0,T];\mathbb R^d)$, see Appendix~\ref{sec Func Ito Calculus}, 
introduced in~\cite{dupire2019functional, cont2010functional, cont2016weak} in the context of functional It\^o calculus. 

The Feynman--Kac theorem provides a probabilistic representation for $F$ and so Monte Carlo methods can be used to approximate $F$ for a single fixed $(t,\omega;\beta)$.
Nevertheless, it is clear that approximating the PPDE solution $F$ across the whole space $[0,T] \times C([0,T];\mathbb R^d) \times B$ is an extremely challenging task. 
A key step is efficiently encoding the information in $\omega$ in some finite dimensional structure. 

Equations of the form~\eqref{eq ppde intro} arise in mathematical finance with $F$ representing a price of some (path-dependent) derivative. 
In this context $\beta$ would be model parameters, $t$ the current time and $\omega$ a path ``stopped at $t$'' representing the price history of some assets. 
Moreover $\nabla_\omega F$ is an object which, in many models, gives access to a ``hedging strategy'' which is of key importance for risk management purposes.
Having an approximation of $F$ that can be quickly evaluated for any $\beta \in B$ is essential for model calibration (see e.g.~\cite{horvath2019deep}).   

\subsection{Main contributions} 
The main contributions in this paper are the following:
\begin{enumerate}[i)]
\item We provide two methods for efficiently encoding the paths $\omega$ using long-short-term-memory (LSTM)-based deep learning methods and path-signatures to approximate the solution of a parabolic PPDE on the whole domain and parameter space. 
\item We provide methods for removing bias in the approximation and a posteriori confidence intervals for then neural network-based approximation. 
\item The algorithms we develop are applicable to parametric families of solutions and hence can be used for efficient model calibration from data.
\item The algorithms we develop provide approximation of $\nabla_\omega F$ thus providing access to the hedging strategies. 
\item We test the algorithms for various models and study their relative performance. The code for our proposed methods and for the numerical experiments can be found at\\ \url{https://github.com/msabvid/Deep-PPDE}. 
\end{enumerate}

\subsection{Overview of existing methods}
Let us now provide a brief overview of other methods available in the literature. 
As has already been mentioned, the probabilistic methods for deep PDE solvers were first explored in~\cite{han2017solving, E_2017} (providing solution for a single point in time and space).
These methods were further extended in~\cite{chanwainam2018machine, pham2019neural, hure2020deep, Beck_2019} to build  deep learning solvers for non-linear PDEs. 
In~\cite{vidales2018unbiased} the probabilistic methods were extended to families of parametric PDEs with approximations across the whole domain.
A different (non-probabilistic) approach was adopted in~\cite{sirignano2017dgm}. 
There a deep neural network is directly trained to satisfy the differential operator, initial and boundary conditions. 
This relies on automatic differentiation to calculate the gradient of the network in terms of its input. 
Related algorithms are being developed in the so-called physics inspired machine learning~\cite{raissi2019physics} where one uses PDEs as regulariser of the neural network. 
See~\cite{e2020algorithms} for a survey of recent efforts to solve PDEs in high dimension using deep learning.

%
%
%
%

Deep Learning methods that approximate the solution of PPDEs have been studied in~\cite{saporito2020pdgm}, where the authors 
extend the work from~\cite{sirignano2017dgm} and they use a LSTM network to include the path-dependency
of the solution of the PPDE. A different approach is proposed in~\cite{jacquier2019deep} where the authors first discretise the space to approximate a PPDE by high-dimensional classical PDE, and use deep BSDE solver approach to solve it. 

Unlike the methods mentioned earlier, the algorighms in this paper can be used for parametric families of solutions of PDEs (PPDEs).
Having approximation of $F(\cdot;\beta)$ for any $\beta \in B$
 allows for swift calibration of models to market data (e.g options prices) since we also have access to $\nabla_{\beta}F$ using automatic differentiation.  
 This is particularly appealing for high dimensional problems or for models for which computation of the pricing operator is costly. 
 This line of research has been recently studied in various settings and with various datasets~\cite{horvath2019deep,Liu_2019,bayer2018deep,stone2019calibrating,hernandez2016model,itkin2019deep,mcghee2018artificial}. 
 Recent work in~\cite{Cuchiero_2020,gierjatowicz2020robust} use Neural SDEs to perform a data-driven model calibration i.e. without using a prior assumption on the form of the dynamics of the price process.

\subsection{Outline} 
In Section~\ref{sec path dependent PDE} we define the notion of Path Dependent PDE, relying
on Functional It\^o Calculus, and we introduce the Feynman--Kac formula extended for path-dependent functionals. 
Section~\ref{sec martingale representation} develops the martingale representation of the discounted price of a path-dependent option, and how it can be used to retrieve the hedging strategy. Section~\ref{sec DL framework} develops the algorithms to approximate the solution of a linear PPDE, built on the probabilistic representation of the solution of the PPDE, and the properties of the conditional expectation and the martingale representation of the discounted price. We finally provide some numerical experiments in Section~\ref{sec results}.  

\subsection{Notation}\label{sec notation} We will use the following notation
\begin{itemize}
\item $t\wedge s = \min(t,s)$
\item Let $m, d, \kappa,p,N \in \mathbb N$, $B\subseteq \mathbb R^p$. Let $F: [0,T] \times C([0,T], \mathbb R^d) \times B \rightarrow \mathbb R^m$ such that $F(\cdot, \cdot; \beta)$ is a non-anticipative functional for all $\beta\in B$ (see Section~\ref{sec path dependent PDE} and Appendix~\ref{sec Func Ito Calculus}). We denote a neural network with weights $\theta\in\mathbb R^{\kappa}$ approximating $F$ as:
\[
\mathcal R_{\theta}[F]: [0,T] \times \mathbb R^N \times B \rightarrow \mathbb R^m
\]
where a path $\omega \in C([0,T], \mathbb R^d)$ is encoded by an element of $\mathbb R^N$ (see Sections~\ref{sec data sim}, \ref{sec short signatures}).
\item $\nabla_{\omega} F(t,\omega; \beta)$ and $\nabla_{\omega}^2 F(t,\omega; \beta)$ are the vector and matrix denoting the first and second order path-derivatives (see Appendix~\ref{sec Func Ito Calculus}) of a non-anticipative functional. Furthermore, the space of non-anticipative functionals admitting time-derivative up to first order and path-derivative up to second order, in addition to satisfying the boundedness property of their time and path-derivatives (Appendix~\ref{sec Func Ito Calculus}) is denoted by $\mathbb C^{1,2}$. 
\item $\text{Sig}^{(n)}_{[t_i, t_j]}$ denotes the path signature
up to the $n$-th iterated integral of a path $(\omega_t)_{t\in[t_i,t_j]}$.
\end{itemize}

\section{PPDE-SDE relationship}\label{sec path dependent PDE}

Appendix~\ref{sec Func Ito Calculus} provides a brief review of the notion of non-anticipative
functionals and their path derivatives. In short, a non-anticipative functional $F: [0,T]\times C([0,T], \mathbb R^d) \rightarrow \mathbb R$ does not look into the future, i.e. given $t\in[0,T]$ and two 
different paths $\omega, \omega'\in C([0,T], \mathbb R^d)$ such that $\omega_{s\wedge t} = \omega'_{s\wedge t}\,\,\, \forall s\in[0,T]$, then $F(t,\omega) = F(t,\omega')$.

The following result allows to represent the solution of a linear PPDE with terminal condition as the expected value of a random variable. 
Fix a probability space $(\Omega, \mathcal F, \mathbb P)$, and consider a continuous process $(X_t)_{t\in[0,T]}$ given by
\begin{equation}\label{eq SDE}
dX_t^{\beta} = b(t,(X_{t\wedge s}^{\beta})_{s\in[0,T]};\beta)dt + \sigma(t,(X_{t\wedge s}^{\beta})_{s\in[0,T]};\beta) dW_t
\end{equation}
where $b, \sigma$ are non-anticipative functionals and $(W_t)_{t\in[0,T]}$
is a Brownian motion, and $\beta\in B\subseteq \mathbb R^p$.
Furthermore, assume that $b, \sigma$ are such that the SDE admits a unique 
strong solution.
On the other hand, let $F$ satisfy the following conditions:
\begin{enumerate}[i)]
	\item it is regular enough admitting time derivative up to first order and path-derivatives up to second order (as defined in Appendix~\ref{sec Func Ito Calculus}).
	\item it is the solution of the following linear PPDE, 
\begin{equation}\label{eq PPDE}
\begin{split}
& \bigg[\partial_t F + b \nabla_\omega F + \frac{1}{2}\text{tr}\left[\nabla_{\omega}^2 F \sigma^*\sigma\right] - rF\bigg](t,\omega;\beta)  = 0\,, \\
& F(T,\omega;\beta)  = g(\omega;\beta)\,,\,\,\, t \in [0,T]\,,\,\, \omega \in C([0,T];\mathbb R^d)\,,\,\,\beta \in B\,.
\end{split}
\end{equation}
\end{enumerate}
Then one can establish a probabilistic representation of $F(t,\omega; \beta)$ via the Feynman-Kac formula. In the following result, we assume $\beta$ fixed, and we abuse the notation to write $F(t,\omega) :=F(t,\omega; \beta)$.

\begin{theorem}[Feynman-Kac formula for path-dependent functionals, see Th. 8.1.13 in ~\cite{bally2016stochastic}]
Consider the functional $g:C([0,T],\mathbb R^d) \rightarrow \mathbb R$,
continuous with respect to the distance $d_{\infty}(\omega,\omega'):=\sup_{t\in[0,T]}|\omega(t)-\omega'(t)|$. If for every 
$(t,\omega)\in \left([0,T],C([0,T],\mathbb R^d)\right)$ the functional $F\in \mathbb C^{1,2}$ verifies~\eqref{eq PPDE},
then $F$ has the probabilistic representation
\begin{equation}\label{eq cond exp}
F(t,\omega) = e^{-r(T-t)}\mathbb E\left[ g((X_t)_{t\in[0,T]}) \bigg | (X_{t\wedge s})_{s\in [0,T]}=(\omega_{t\wedge s})_{s\in [0,T]}\right].	
\end{equation}
with $(X_t)_{t\in[0,T]}$ given by~\eqref{eq SDE}. 
\end{theorem}

A direct consequence of the Feynman--Kac formula for path-dependent functionals is that 
solving~\eqref{eq PPDE} is equivalent to pricing the path-dependent option with payoff at $T$ given by $g((X_s)_{s\in[0,T]})\in L^2(\mathcal F_T)$ where $(X_t)_{t\geq 0}$ is the solution of~\eqref{eq SDE} and $(\mathcal F_t)_{t\geq 0}$ denotes the filtration generated by $(X_t)_{t\geq 0}$.
We will build two algorithms based on two different approaches:
\begin{enumerate}[i)]
\item Option pricing using the martingale representation theorem (Algorithm~\ref{alg martingale representation}).
\item Considering the conditional expectation in ~\eqref{eq cond exp} as the orthogonal projection of
$g(X_T)$ on $\mathcal L^2(\mathcal F_t)$ where $(\mathcal F_t)_{t\geq 0}$ is the filtration generated by $(X_t)_{t\geq 0}$ (Algorithm~\ref{alg orthogonal projection}). 
\end{enumerate}

\section{Option pricing via Martingale Representation Theorem}\label{sec martingale representation}

In this section we assume constant interest rate and a complete market but the results readily extend to the case of stochastic interest rates and incomplete markets.

Let $(\Omega, \mathcal F, \mathbb Q)$ be a probability space where $\mathbb Q$ is the risk-neutral measure. 
Consider an $\mathbb R^{d}$-valued Wiener process $W=(W^j)_{j=1}^{d} = ((W^j_t)_{t\geq 0})_{j=1}^{d}$.
We will use $(\mathcal F^W_t)_{t\geq 0}$ to denote the filtration generated by $W$. 
Consider an $D\subseteq \mathbb R^d$-valued, continuous, stochastic process $X=(X^i)_{i=1}^d = ((X^i_t)_{t\geq 0})_{i=1}^d$ that is adapted to $(\mathcal F^W_t)_{t\geq 0}$.
We will use $(\mathcal F_t)_{t\geq 0}$ to denote the filtration generated by $X$. 

We recall that we denote by $\beta\in B\subseteq\mathbb R^p$ the family of parameters of the dynamics of the underlying asset (for instance, in the Black--Scholes model with fixed risk-free rate, $\beta$ denotes the volatility).
Let $g : C([0,T],\mathbb R^d) \times B \to \mathbb R$ such that $g(\cdot, \beta)$ is a measurable function for each $\beta \in B\subseteq \mathbb R^p$. 
We shall consider contingent claims of the form $g((X_s^{\beta})_{s\in[0,T]}; \beta)$.
This means that we can consider path-dependent derivatives.
Finally, let $r$ be some risk-free rate, and consider, for each $\beta$, the SDE 
\[
dX_t^{\beta} = rX_t^{\beta}dt + \sigma(t, (X_{s\wedge t}^{\beta})_{s\in[0,T]}; \beta)dW_t.
\]
We immediately see that $\bar X_t^{\beta} = (e^{-rt}X_t^{\beta})_{t\in[0,T]}$ is a (local) martingale. 


In order to price an option at time $t$ with payoff $g$, under appropriate assumptions on $g$ and $\sigma$, the random variable
\[
M_t^{\beta} := \mathbb E\left[ e^{-rT}g((X_s^{\beta})_{s\in[0,T]}; \beta)\bigg | \mathcal F_t^{\beta}\right]
\]
is square-integrable. $M_T^{\beta} = e^{-rT}g((X_s^{\beta})_{s\in[0,T]})$ is the discounted payoff at $T$. Hence $F(t, (X_{t\wedge s}^{\beta})_{s\in[0,T]}; \beta) = e^{rt}M_t^{\beta}$ is the fair price of the option  with payoff $g$ at time $t$. By the Martingale representation theorem, for each $\beta$ there exists a unique $\mathcal F_t$-adapted process $Z_s^{\beta}$
with $\mathbb E\left[\int_0^T (Z_s^{\beta})^2 ds \right]<\infty$ such that
\begin{equation}\label{eq martingale repr}
M_T^{\beta} = \mathbb E[ M_T^{\beta} | \mathcal F_0] + \int_0^T Z_s^{\beta} dW_s.
\end{equation}
In order to retrieve the real hedging strategy from the martingale representation, it is necessary to apply the Itô formula for non-anticipative functionals 
of a continuous semimartingale, see~\cite{dupire2019functional,Cont_2013}:
\begin{proposition}. Let $X$ be a continuous semimartingale defined on $(\Omega, \mathcal F, \mathbb Q)$. Then, for any non-anticipative functional $F\in \mathbb C^{(1,2)}$ (introduced in Section~\ref{sec notation}) and any $t\in[0,T]$, we have 
\begin{equation}\label{eq functional ito formula}
\begin{split}
dF(t,(X_{s\wedge t})_{s\in[0,T]}) = & \partial_t F(t,(X_{s\wedge t})_{s\in[0,T]})dt + \nabla_{\omega} F(t,(X_{s\wedge t})_{s\in[0,T]})dX_t \\
 & + \frac{1}{2}\text{tr}\left((\nabla_{\omega}^2 F(t,(X_{s\wedge t})_{s\in[0,T]}))dX_tdX_t\right).
\end{split}
\end{equation}
\end{proposition}

Let $\bar F_t := e^{-rt}F(t, (X_{t\wedge s}^{\beta})_{s\in[0,T]}; \beta)$ be the discounted 
price of the option with payoff $g$ at time $t$. Then, using~\eqref{eq functional ito formula},
\begin{equation}\label{eq ito discounted price}
d\bar F_t = \left(-rF + \partial_t F  + \frac{1}{2}\text{tr}(\nabla_{\omega}^2 F \sigma^* \sigma) + r X_t \nabla_{\omega} F\right)e^{-rt}dt + \nabla_{\omega} F \cdot e^{-rt}\sigma  dW_t\,.
\end{equation}
Moreover, $\bar F_t$ is the value of the discounted portfolio at $t$ (since the market is complete) thus the coefficient of $dt$ in~\eqref{eq ito discounted price} is 0. Noting that $d\bar X_t^{\beta} = e^{-rt}\sigma(t, (X_{t\wedge s}^{\beta})_{s\in[0,T]}; \beta) dW_t$, we get
\[
d\bar F_t = \nabla_{\omega} F d\bar X_t.
\]   
Hence after replacing in~\eqref{eq martingale repr} one can retrieve the hedging strategy
\begin{equation}\label{eq martingale repr 2}
M_t^{\beta} = \mathbb E[M_t^{\beta}|\mathcal F_0] + \int_0^t \nabla_{\omega} F\, d\bar X_s^{\beta}, \quad M_T^{\beta} = e^{-rT}g((X_s^{\beta})_{s\in[0,T]}; \beta)\,.
\end{equation}
Both $M_t^{\beta}$ and the stochastic integral in~\eqref{eq martingale repr 2} are martingales. This will be used in Section~\ref{sec learning martingale representation} to build a learning task to solve the BSDE~\eqref{eq martingale repr} to jointly approximate the fair price of the option $F$, and the hedging strategy $\nabla_{\omega} F$.

\section{Deep PPDE solver Methodology}
\label{sec DL framework}
In this section we present the PPDE solver methodology consisting on the data simulation scheme (Section~\ref{sec data sim}). We briefly define the signature of a path, that we will use as a path feature extractor (Section~\ref{sec short signatures}). We describe two optimisation tasks to approximate the price 
of path-dependent derivatives, relying on conditional expectation properties (Section~\ref{sec orthogonal projection}), and on the martingale representation of the discounted price (Section~\ref{sec learning martingale representation}). We present the learning scheme
 leveraging the learning methods, the different deep network architectures considered, and the path signatures (Section~\ref{sec DL solver networks}).  Finally, we present the evaluation metrics used in the numerical experiments (Section~\ref{sec evaluation scheme}).

%

\subsection{Data simulation scheme}\label{sec data sim}
We consider the SDE with path-dependent coefficients
\begin{equation}\label{eq sde}
dX_t^{\beta} = rX_t^{\beta}dt + \sigma((X_{t\wedge s}^{\beta})_{s\in[0,T]}; \beta) dW_t,\,\,\,t\in[0,T]\,,\,\,\, X(0) = x_0,
\end{equation}
and the path-dependent payoff,
$g : C([0,T],\mathbb R^d)\times B \to \mathbb R$. We will consider two different time discretisations. 
\begin{enumerate}[i)]
\item First, a 
fine time discretisation $\pi^f:=\{0=t_0^f<t_1^f<\ldots<t_N^f=T\}$ used by the numerical SDE solver to sample paths from~\eqref{eq sde}.
\item We consider a second, coarser, time discretisation $\pi^c:=\{0=t_{0}^c<t_1^c<\ldots<t_N^c=T\} \subseteq \pi^f$ on which we learn the deep learning approximation of the price of the option.   
\end{enumerate}
Furthermore, we fix the distribution of $\beta\in B\subseteq \mathbb R^p$.
We will denote the discretisation of $(X_t^{\beta})_{t\in[0,T]}$ in $\pi^f$ using Euler scheme as $(X^{\beta,\pi^f}_{t})_{t\in\pi^f}$

\subsection{Path signatures as feature extractors of paths}\label{sec short signatures}
The input to $\mathcal R_{\theta}[F]$ is a discretisation of elements of $[0,T] \times C([0,T], \mathbb R^d) \times B$. given by the numerical approximation from the SDE solver, $(x_{t}^{\beta,\pi^f})_{t\in\pi^f} \in \mathbb R^N$. If $\pi^f$ is a fine partition of $[0,T]$, then the input to $\mathcal R_{\theta}[F]$ will be high dimensional. We explore two alternatives to avoid feeding the whole path to the neural network. The first naive approach is feeding the path generated with the SDE solver on $\pi^f$ but evaluated on the coarser time discretisation $(x_{t}^{\beta,\pi^f})_{t\in\pi^c}$. In such approach the information carried by the path in the the points of $\pi^f$ that are not in $\pi^c$ is lost. Alternatively, we use path signatures to capture a description of the path on $\pi^f$, and provide it as an input to the neural network. 

We refer the reader to Appendix~\ref{sec path signatures} and the references therein for supplementary material on path signatures. Let $T((\mathbb R^d)) := \bigoplus_{k=0}^{\infty}(\mathbb R^d)^{\otimes k}$ be a tensor algebra space. Then, the signature of $X:[a,b]\rightarrow \mathbb R^d$ is an element of the tensor algebra $T((\mathbb R^d))$,
\[
\text{Sig}_{a,b}(X) =  (1,S(X)_{a,b}^{(1)}, S(X)_{a,b}^{(2)},\ldots)\in T((\mathbb R^d)).
\]
where
\[
S(X)_{a,t}^{(k)} = \int_{a<t_1<t_2<\ldots<t_k<t} dX_{t_1} \otimes \ldots \otimes dX_{t_l} \in T^k(\mathbb R^d)
\]
In short, the signature of a path determines the path essentially uniquely (up to time reparametrisations), providing top-down description of the path: low order terms of the signature capture global properties of the path (for instance $S(X)_{a,b}^{(1)}$ provides the change of each of the path coordinates between $a$ and $b$), whereas higher order terms give information on the local structure of the path.          

Let $(x_t^{\beta, \pi^f})_{t\in\pi^f}$ be the discretisation of a path on $\pi^f$ generated by the SDE solver. Then we encode it as \textit{stream of signatures} as follows
\begin{equation}\label{eq path encoding}
(y_{t_k}^{\beta,\pi^f})_{t_{k}\in\pi^c} := \left(\text{Sig}_{[t_{k},t_{k+1}]}^{(n)}(x_t^{\beta,\pi^f})_{t\in\pi^f}\right)_{t_k\in\pi^c},	
\end{equation}
i.e. such that each element of $(y_{t_k}^{\beta,\pi^f})_{t_{k}\in\pi^c}$ is the signature of $(x_t^{\beta, \pi^f})_{t\in\pi^f}$ between consecutive steps of $\pi^c$.

\subsection{Learning conditional expectation as orthogonal projection}
\label{sec orthogonal projection}
The following theorem recalls a well known property of conditional expectations:
\begin{theorem}\label{th conditional}
Let $X\in \mathcal L^2(\mathcal F)$. Let $\mathcal G \subset \mathcal F$ be a sub $\sigma$-algebra. There exists a random variable $Y\in \mathcal L^2(\mathcal G)$ such that 
\begin{equation}\label{eq orthogonal projection}
\mathbb E[| X - Y |^2] = \inf_{\eta \in \mathcal L^2(\mathcal G)} \mathbb E[| X- \eta  |^2].	
\end{equation}
The minimiser, $Y$, is unique and is given  by $Y=\mathbb E[X | \mathcal G]$.
\end{theorem}
The theorem tells us that the conditional expectation is an orthogonal
projection of a random variable $X$ onto $\mathcal L^2(\mathcal G)$. 
To formulate the learning task in our problem, we replace $X$ in~\eqref{eq orthogonal projection} by $e^{-r(T-t)}g((X_s^{\beta})_{s\in[0,T]}; \beta)$. We also replace $\mathcal G$ by $\mathcal F^{\beta}_t$,  and $\mathcal F$ by $\mathcal F_T^{\beta}$.
Then by Theorem~\ref{th conditional} 
\[
\begin{split}
F(t, (X_{s}^{\beta})_{s\in[0,T]}; \beta) & =  \mathbb E[e^{-r(T-t)}g((X^{\beta}_s)_{s\in[0,T]}; \beta)|\mathcal F_t^{\beta}]  \\\
& =  \arg\inf_{\eta \in\mathcal L^2\left(\mathcal F_t\right)} \mathbb E[|e^{-r(T-t)}g((X_s^{\beta})_{s\in[0,T]}; \beta) - \eta|^2]
\end{split}
\]
By the Doob--Dynkin Lemma~\cite[Th. 1.3.12]{cohen2015stochastic} 
we know that every $\eta \in L^2(\mathcal F_t^{\beta})$ can be expressed as 
$\eta = h_t((X^{\beta}_{s\wedge t})_{s\in[0,T]})$ for some appropriate measurable $h_t$.
For the practical algorithm we restrict the search for the function $h_t$ to the class that can be expressed as deep neural networks. 

We then consider deep network approximations of the price of the
path-dependent option at any time $t_k^c$ in the time partition $\pi^c$, either by directly using the path evaluated at each time step of $\pi^c$ or by using the stream of signatures~\eqref{eq path encoding}:   
\begin{equation}\label{eq possible inputs}
\text{i)} \,\,(\mathcal X_{t}^{\beta, \pi^f})_{t\in\pi^c} :=  (x_{t}^{\beta, \pi^f})_{t\in\pi^c} \qquad \text{or ii)}\,\, (\mathcal X_{t}^{\beta, \pi^f})_{t\in\pi^c} := (y_{t}^{\beta, \pi^f})_{t\in\pi^c}  
\end{equation}
and set the learning task as 
\begin{equation}
\label{eq learning task orthogonal projection}
\theta^* = 	\argmin_{\theta} \mathbb E_{\beta} \left[\mathbb E_{(X_t^{\beta, \pi^f})_{t\in\pi^f}} \left[\sum_{k=0}^N \left(e^{-r(T-t_k)}
g((X_t^{\beta, \pi^f})_{t\in \pi^f}; \beta) - \mathcal R_{\theta}[F](t_k, (\mathcal X_{t}^{\beta, \pi^f})_{t\in\pi^c}, \beta) \right)^2\right]\right]\,.
\end{equation}

We observe that the inner expectation in~\eqref{eq learning task orthogonal projection} is taken across all paths generated using the numerical SDE solver on~\eqref{eq sde} on $\pi^f$ for a fixed $\beta$ and it allows to price an option for such $\beta$. 
The outer expectation is taken on $\beta$ for which the distribution is fixed (as specified in the data simulation scheme), thus allowing to find the optimal neural network weights $\theta^*$ to price the parametric family of options.

\begin{algorithm}[H]
\caption{Data simulation}	
\label{alg data}
\begin{algorithmic}
\STATE{Initialisation: $N_{\text{trn}}\in \mathbb N$ large}, distribution of $\beta$.
\FOR{$i:1:N_{\text{trn}}$}
\STATE{ Generate training paths 
$(x_{t}^{\beta, \pi^f, i})_{t \in \pi^f}$ using the numerical SDE solver on ~\eqref{eq sde} and sampling from the distribution of $\beta$ .
}
\ENDFOR
\IF{Network input is path discretisation at $\pi^c$}
\STATE{$(\mathcal X_t^{\beta,\pi^f, i})_{t\in\pi^c}$ := $(x_t^{\beta,\pi^f, i})_{t\in\pi^c}$}
\ELSE
\STATE{$(\mathcal X_{t_k}^{\beta,\pi^f, i})_{t_{k}\in\pi^c} := \left(\text{Sig}_{[t_{k},t_{k+1}]}^{(n)}(X_t^{\beta,\pi^f, i})_{t\in\pi^f}\right)_{t_k\in\pi^c}$ i.e. $\mathcal X$ is the process of signatures of the path between consecutive time-points of $\pi^c$.
}
\ENDIF
\RETURN $(x_t^{\beta,\pi^f, i})_{t\in\pi^c}, (\mathcal X_{t_k}^{\beta,\pi^f, i})_{t_{k}\in\pi^c}$ for $i=1,\ldots,N_{\text{trn}}$.
\end{algorithmic}
\end{algorithm}

\begin{algorithm}[H]
\caption{Learning orthogonal projection}	
\label{alg orthogonal projection}
\begin{algorithmic}
\STATE{Initialisation: Weights $\theta$ of networks $\mathcal R[F]_{\theta}$, $N_{\text{trn}}\in \mathbb N$ large}, distribution of $\beta$.
\STATE{Use SGD to find $\theta^{*}$, where in each iteration of SGD we generate a batch of paths (or a batch of stream of signatures) using Algorithm~\ref{alg data}.   
 \[
 \theta^{*}= {\argmin_{\theta}}\,\,\, \mathbb E^{\mathbb Q^{N_{\text{trn}}}} \left[ \sum_{k=0}^{N_{\text{steps}}-1}\left(e^{-r(T-t_k)}
g((X_{t}^{\beta,\pi^f})_{t\in \pi^f})  - \mathcal R[F]_{\theta}(t_k,(\mathcal X_{t}^{\beta,\pi^f})_{t\in \pi^c}, \beta) \right)^2\right]
 \]
 where $\mathbb E^{\mathbb Q^{N_{\text{trn}}}}$ denotes the empirical mean. 
 }
\RETURN $\theta^{*}$.
\end{algorithmic}
\end{algorithm}


\subsection{Learning martingale representation of the option price}\label{sec learning martingale representation}
From Section~\ref{sec martingale representation}, for fixed $\beta$ the discounted price of the option with payoff $g$
is given by 
\begin{equation}
M_T^{\beta} = \mathbb E[M_T^{\beta}|\mathcal F_0] + \int_0^T \nabla_{\omega} F\, d\bar X_s^{\beta}.
\end{equation}
Since both $(M_t^{\beta})_{t\in[0,T]}$ and the stochastic integral are martingales, after taking expectations conditioned on $\mathcal F_s^{\beta}, \mathcal F_t^{\beta}$ on both sides for $s<t\leq T$ one gets
\begin{equation}\label{eq learning task martingale}
M_t^{\beta} = M_s^{\beta} + \int_s^t \nabla_{\omega} F \,d\bar X_s^{\beta}.
\end{equation}
After replacing the pair $s, t$ in~\eqref{eq learning task martingale} by all possible consecutive times in $\pi^c$, 
 one obtains a \textit{backward} system of equations starting from the final condition $M_T^{\beta} = e^{-rT}g((X_{t}^{\beta})_{t\in \pi^f};\beta)$ 
\begin{equation}\label{eq system BSDE}
M_{t_{k}}^{\beta} =  M_{t_{k-1}}^{\beta} + \int_{t_{k-1}}^{t_k} \nabla_{\omega} F\, d\bar X_s^{\beta}, \, \, \, \, k=N,\ldots,1.
\end{equation}
Considering the deep learning approximations of the price of the option and the hedging strategy with two neural networks whose input is either the path evaluated on $\pi^c$ or the stream of signatures~\eqref{eq path encoding},
and replacing them in~\eqref{eq system BSDE} then the sum of the $L^2$-errors that arise from~\eqref{eq system BSDE} yields the optimisation task to learn the weights of $\mathcal R[F]_{\theta}, \mathcal R[\nabla_{\omega} F]_{\phi}$: 
\begin{equation} \label{eq loss function BSDE solver}
\begin{split}
 (\theta^*, \phi^*) := & \argmin_{(\theta,\phi)} \mathbb E_{\beta} \bigg[\mathbb E_{(X_t^{\beta, \pi^f})_{t\in\pi^f}} 
 \bigg[ \left(g((X_{t}^{\beta, \pi^f})_{t\in\pi_f}; \beta) - \mathcal R[F]_{\theta}(t_N,(\mathcal X_t^{\beta, \pi^f})_{t\in\pi^c}; \beta)\right)^2 + \\
 & \qquad \qquad \qquad \qquad \qquad \qquad \sum_{m=0}^{N-1}|\mathcal E^{(\theta,\phi)}_{m+1}|^2 \bigg]\bigg]\,,\\
\mathcal E^{(\eta,\theta)}_{m+1} 
:= & e^{-rt_{m+1}}\mathcal{R}[F]_{\theta}(t_{m+1}, (\mathcal X_{t}^{\beta,\pi^f})_{t\in\pi^c}; \beta) -   
e^{-rt_{m}}\mathcal{R}[F]_{\theta}(t_m, (\mathcal X_{t}^{\beta, \pi^f})_{t\in\pi^c}; \beta) \\
& \quad  - e^{-rt_{m}} \mathcal{R}[\nabla_{\omega} F]_{\phi}(t_{m}, (\mathcal X_{t}^{\beta, \pi^f})_{t\in\pi^c}; \beta)\sigma(t_{m},(\mathcal X_{t}^{\beta, \pi^f})_{t\in\pi^c}; \beta) \Delta W_{t_{m}}\,,
\end{split}
\end{equation}
where as before $(\mathcal X_{t}^{\beta, \pi^f})_{t\in\pi^c}$ denotes the choice of the input used in the learning algorithm. 

Learning task in pseudocode is provided in Algorithm~\ref{alg martingale representation}. 

\begin{algorithm}[H]
\caption{Learning Martingale representation}
\label{alg martingale representation}
\begin{algorithmic}
\STATE{Initialisation: Weights $\theta, \phi$ of networks $\mathcal R[F]_{\theta}, \mathcal R[\nabla_{\omega} F]_{\phi}$, $N_{\text{trn}}\in \mathbb N$ large, distribution of $\beta$.}
\STATE{
Use SGD to find $(\theta^{*},\phi^*)$, where in each iteration of SGD we generate a batch of paths (or a batch of stream of signatures) using Algorithm~\ref{alg data}.   
\[
\begin{split}
 (\theta^*, \phi^*) := & \argmin_{(\theta,\phi)} \mathbb E^{\mathbb Q^{N_{\text{trn}}}} 
 \bigg[ \left(g((X_{t}^{\beta,\pi^f})_{t\in\pi_f}; \beta) - \mathcal R[F]_{\theta}(t_N,(\mathcal X_t^{\beta,\pi^f})_{t\in\pi^c}; \beta)\right)^2 \\
& \qquad \qquad + \sum_{m=0}^{N-1}|\mathcal E^{(\theta,\phi)}_{m+1}|^2 \bigg]\,,\\
\mathcal E^{(\eta,\theta)}_{m+1} 
:= & e^{-rt_{m+1}}\mathcal{R}[F]_{\theta}(t_{m+1}, (\mathcal X_{t}^{\beta,\pi^f})_{t\in\pi^c}; \beta) -   
e^{-rt_{m}^c}\mathcal{R}[F]_{\theta}(t_m, (\mathcal X_{t}^{\beta, \pi^f})_{t\in\pi^c}; \beta) \\
& \quad  - e^{-rt_{m}} \mathcal{R}[\nabla_{\omega} F]_{\phi}(t_{m}, (\mathcal X_{t}^{\beta, \pi^f})_{t\in\pi^c}; \beta)\sigma(t_{m},(\mathcal X_{t}^{\beta, \pi^f})_{t\in\pi^c}; \beta) \Delta W_{t_{m}}\,,
\end{split}
\]
where $\mathbb E^{\mathbb Q^{N_{\text{trn}}}}$ denotes the empirical mean. 
}
\RETURN $(\theta^{*}, \phi^*)$.
\end{algorithmic}
\end{algorithm}


\subsection{Unbiased PPDE solver}\label{sec unbiased} An unbiased estimator of the solution of the PPDE at any $t\in[0,T]$ can be obtained with a Monte Carlo estimator using the Feynman--Kac theorem. 
We will use notation from Section~\ref{sec martingale representation} and fix $\beta \in B$ as the parameters and $t\in [0,T]$ as the current time. 
From~\eqref{eq martingale repr 2} we see that with $M_t = e^{-rt} F(t,(X^\beta_{t\wedge s})_{s\in[0,T]};\beta)$ we have
\begin{equation*}
\label{eq for cv sec 1}	
M_t = M_T - \int_t^T e^{-rs'}\big[(\nabla_\omega F)  \sigma\big](s',(s', (X^\beta_{s'\wedge s})_{s\in[0,T]};\beta)\,dW_{s'}\,.
\end{equation*}
If we replace the exact gradient $\nabla_\omega F$ by $\mathcal R_\phi[\nabla_\omega F]$ we can use this as control variate. Let
\begin{equation*}
\label{eq for cv sec 2}		
M_t^{\text{cv},\phi}:= M_T - \int_t^T e^{-rs'}\big(\mathcal R_\phi[\nabla_\omega F]  \sigma\big)(s',(s', (X^\beta_{s'\wedge s})_{s\in[0,T]};\beta)\,dW_{s'}\,.
\end{equation*}
While $\mathbb V\text{ar}[M_t] = 0$ for a good approximation $\mathcal R_\phi[\nabla_\omega F]$
to $\nabla_\omega F$ we will have $\mathbb V\text{ar}[M^{\text{cv}
,\phi}_t]$ small. 
Consider $(W^{i})_{i=1}^N$, $N$ i.i.d copies of $W$.
Let $\mathbb Q^{N} := \frac{1}{N}\sum_{i=1}^N \delta_{X^{i}}$ be the empirical approximation of the risk neutral measure.
Then 
\begin{equation}
\label{eq for cv sec 3}
F(t,(X^\beta_{t\wedge s})_{s\in[0,T]};\beta) \approx e^{rt}\mathbb E^{\mathbb Q^N} \big[ M_t^{\text{cv},\phi} | \mathcal F_t \big]\,.	
\end{equation}
Central Limit Theorem tells us that
\[
\mathbb Q\left(F(t,(X^\beta_{t\wedge s})_{s\in[0,T]};\beta) \in \left[e^{rt}\,  \mathbb E^{\mathbb Q^{N}} \left[M^{\text{cv},\phi}_t | \mathcal F_t \right] \pm z_{\alpha/2}\frac{\Sigma}{\sqrt{N}}\right] \right) \to 1 \quad \text{as} \quad N\to \infty\,.
\]
Here $\Sigma^2 := \mathbb V\text{ar}[e^{rt}M^{\text{cv}
,\phi}_t]$ is small by construction and $z_{\alpha/2}$ is such that $1-\text{CDF}_Z(z_{\alpha/2}) = \alpha/2$ with $Z$ the standard normal distribution.
Hence~\eqref{eq for cv sec 3} is a very accurate approximation even for small values of $N$ since $\Sigma^2 = \mathbb V\text{ar}[e^{rt}M^{\text{cv}
,\phi}_t]$ is small by construction.

An unbiased approximation of $F(t,(X^\beta_{t\wedge s})_{s\in[0,T]};\beta)$ together with confidence intervals can be obtained using Algorithm~\ref{alg unbiased PPDE solver}. 

%
%

\begin{algorithm}
\caption{Unbiased PPDE solver}
\label{alg unbiased PPDE solver}
\begin{algorithmic}
\STATE{Input: current time $t$, path history $(x_{t\wedge s}^{\pi^f})_{s\in\pi^f}$, model parameters $\beta$, confidence level $\alpha$,  $N_{MC}\in\mathbb N$, optimal weights $\phi^*$ for $\mathcal R[\nabla_{\omega} F]_{\phi^*}$.
}
\STATE{
Use Algorithm~\ref{alg data} to generate $N_{MC}$ paths using the numerical SDE solver on~\eqref{eq sde} \textbf{starting from} $(t,x_t^{\beta, \pi^f})$, obtaining
\[
(x_s^{\beta,\pi^f, i})_{s\in\pi^c}, (\mathcal X_{s}^{\beta,\pi^f, i})_{s\in\pi^c} \,\,\text{for}\,\, i=1,\ldots,N_{MC}
\]
such that for each $i$, $(x_{t\wedge s}^{\beta,\pi^f, i})_{s\in\pi^c} = (x_{t\wedge s}^{\beta, \pi^f})_{s\in\pi^c}$.
}
\STATE{
Use the generated $N_{MC}$ paths to calculate 
\[
\begin{split}
F^{\text{MC}}(t,(x_{t\wedge s}^{\beta, \pi^f})_{s\in\pi^c}; \beta) = & \,\,\mathbb E^{\mathbb Q^{N_{MC}}}\left[ e^{rt} M_t^{cv, \phi^*}\, \big |\, (X^\beta_{t\wedge s})_{s\in[0,T]} = (x_{t\wedge s}^{\beta, \pi^f})_{s\in\pi^c}\right] \\
(\Sigma^{\text{MC}})^2 := & \,\, \mathbb V\text{ar}^{\mathbb Q^{N_{MC}}} \left[e^{rt}M_t^{cv, \phi^*}\, \big |\, (X^\beta_{t\wedge s})_{s\in[0,T]} = (x_{t\wedge s}^{\beta, \pi^f})_{s\in\pi^c} \right]
 \end{split}
\]
where 
\[
M_t^{cv, \phi^*} := e^{-rT}g((X_s^{\beta, \pi^f})_{s\in\pi^f}, \beta) - \sum_{s\in \pi^c, s\geq t} e^{-rs} (\mathcal R[\nabla_{\omega}F]_{\phi^*}\sigma)(s, (\mathcal X_s^{\beta, \pi^f})_{s\in \pi^c})\Delta W_s, 
\]
and $\mathbb E^{\mathbb Q^{N_{MC}}},\mathbb V\text{ar}^{\mathbb Q^{N_{MC}}}$ denote the empirical mean and the empirical variance of the Monte Carlo estimator.
}
\STATE{
Calculate the confidence interval of the unbiased estimator:
\[
I := \left(F^\text{MC}(t,(x_{t\wedge s}^{\beta, \pi^f})_{s\in\pi^c}; \beta) \pm z_{\alpha/2}\frac{\Sigma^{\text{MC}}}{\sqrt{N_{MC}}}\right).
\]
}

\RETURN de-biased estimate $F^\text{MC}(t,(x_{t\wedge s}^{\beta, \pi^f})_{s\in\pi^c}; \beta)$ and confidence interval $I$.
\end{algorithmic}
\end{algorithm}

\subsection{Network architectures: LSTM and Feed Forward Networks}\label{sec DL solver networks}
We explore using LSTM networks and Feed Forward Networks (Appendix~\ref{sec network architecture}) for the parameterisation of $F, \nabla_{\omega} F$
in algorithms~\ref{alg orthogonal projection},~\ref{alg martingale representation}.
\subsubsection{Feedforward networks}
Let $\mathcal R[F]_{\theta}$ be a feedforward network. Since it needs to be a non-anticipative functional, then we will train it using 
\[
\mathcal R[F]_{\theta}(t_k, (\mathcal X_{t}^{\beta, \pi^f})_{t\in \pi^c}, \beta) := \mathcal R[F]_{\theta}(t_k, (\mathcal X_{t_k\wedge t}^{\beta, \pi^f})_{t\in \pi^c}, \beta)
\]
to make $\mathcal R[F]_{\theta}$ non-anticipative. 
If the input is the stream of signatures~\eqref{eq path encoding}, then we abuse the notation for 
\[
(\mathcal X_{t_m\wedge t_k}^{\beta,\pi^f})_{t_k\in\pi^c} := \left(\text{Sig}_{[t_{k},t_{k+1}]}^{(n)}(x_{t\wedge t_m}^{\beta,\pi^f})_{t\in\pi^f}\right)_{t_k\in\pi^c}
\]
i.e. the \textit{stopped stream of signatures} at $t_m$ is the stream of signatures of the path $(x_{t\wedge t_m}^{\beta,\pi^f})_{t\in\pi^c}$ stopped at $t_m$. 

\begin{figure}
\centering
  \includegraphics[width=0.8\linewidth]{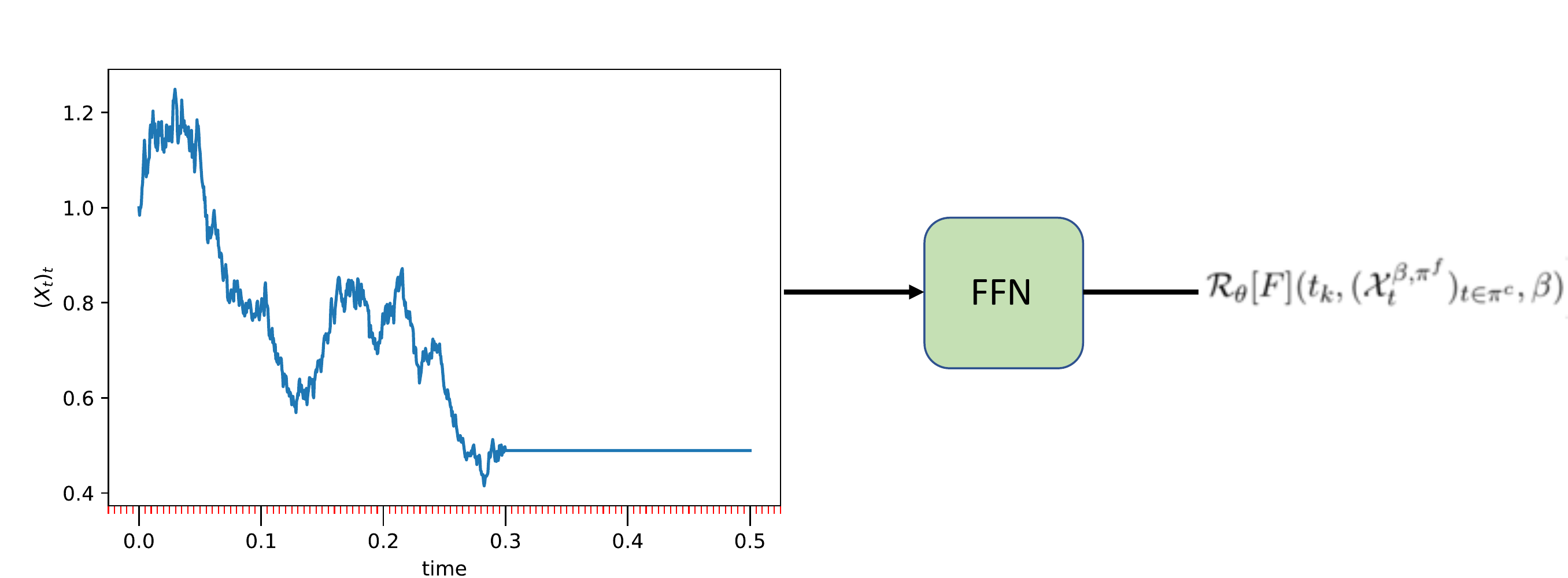}
\caption{Diagram FFN network using stopped path as input}
\label{fig FFN path diagram}
\end{figure}

\subsubsection{LSTM networks}
\begin{figure}
\centering
  \includegraphics[width=0.5\linewidth]{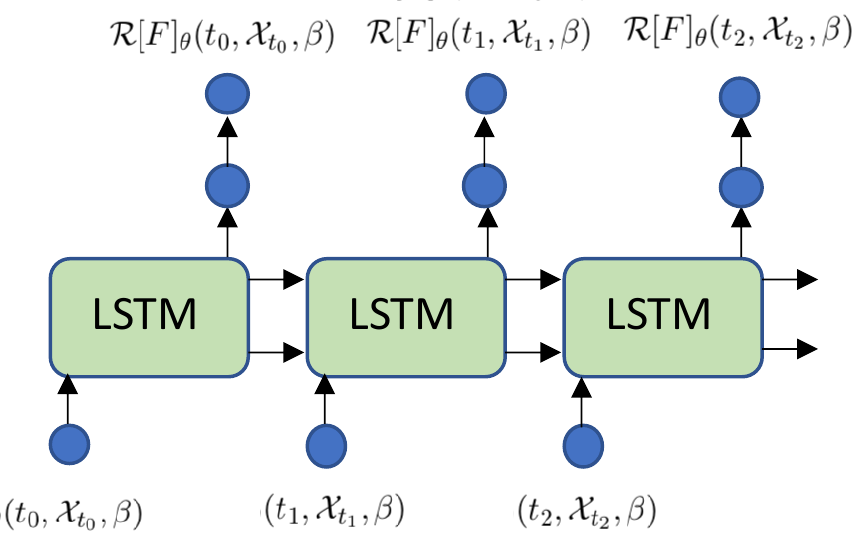}
\caption{Diagram of LSTM network using path as input}
\label{fig LSTM path diagram}
\end{figure}

\begin{figure}
\centering
  \includegraphics[width=0.6\linewidth]{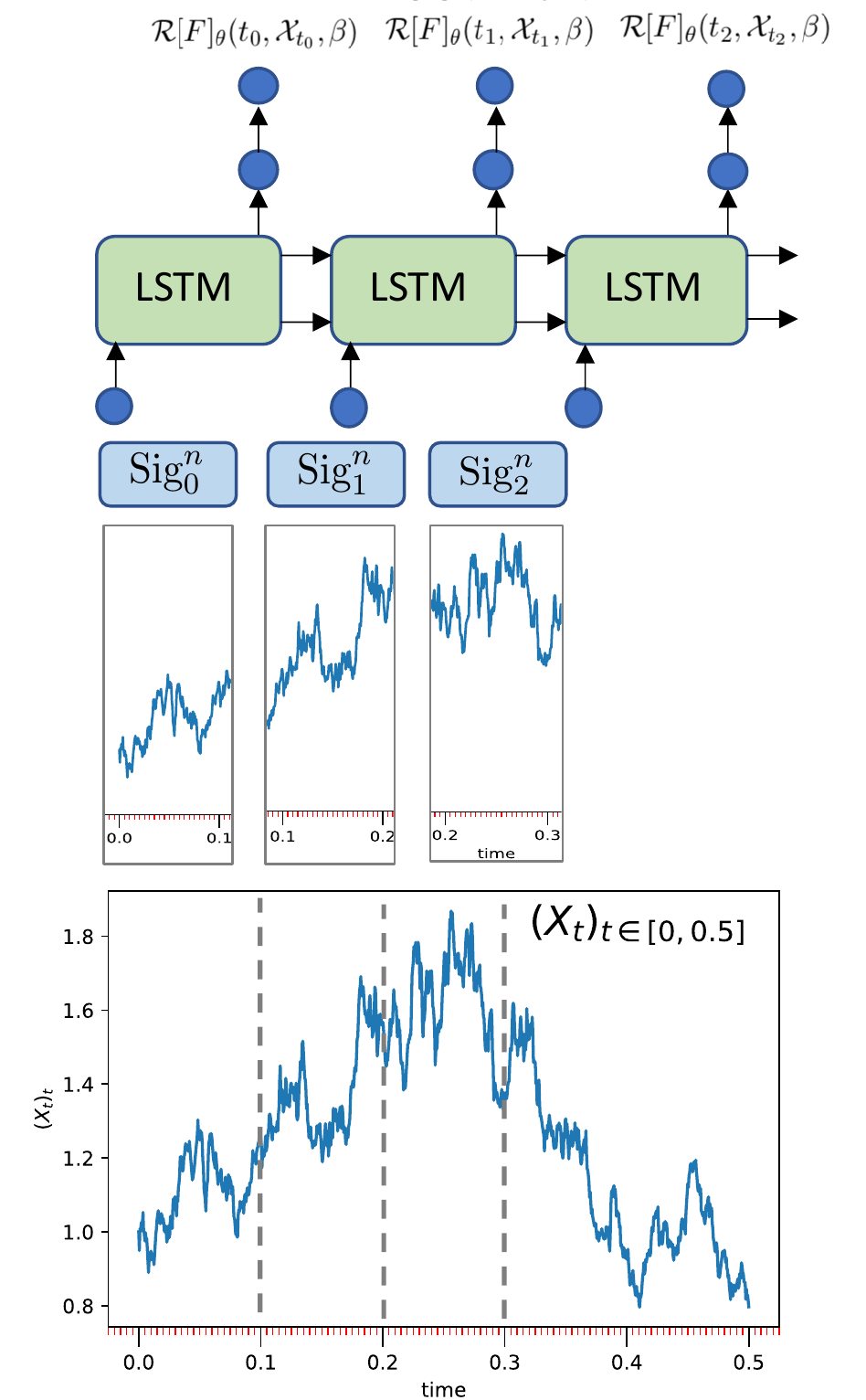}
\caption{Diagram of LSTM network using path signature as input}
\label{fig LSTM sign diagram}
\end{figure}

Recurrent neural networks are a more natural approach to parametrise non-anticipative functionals, since their sequential output is adapted to the input, in the sense that $\mathcal{R}[F]_{\theta}(t_k, (x_{t}^{\beta, \pi^f})_{t\in\pi^c})$ is built without looking into the future of the path at $t_k$ (Figure~\ref{fig LSTM path diagram}). 
Figure~\ref{fig LSTM sign diagram} displays the deep learning setting in the particular case where we use the stream of signatures~\eqref{eq path encoding} as an input to the LSTM network.

\subsection{Evaluation scheme}\label{sec evaluation scheme}
In this section we provide the evaluation measures of the deep solvers introduced in Algorithms~\ref{alg orthogonal projection} and~\ref{alg martingale representation}. 
\subsubsection{Integral error of the option price parametrisation}
We provide the absolute integral error of the solution of the path-dependent PDE,
calculated on a test set for a fixed $\beta$. Recall that the Feynman--Kac formula tells us that the solution of the PPDE has a stochastic representation
\[
F(t,\omega; \beta) = e^{-r(T-t)}\mathbb E\left[ g((X_t^{\beta})_{t\in[0,T]}; \beta) | (X_{t\wedge s}^{\beta})_{s\in [0,T]}=(\omega_{t\wedge s}^{\beta})_{s\in [0,T]}\right].	
\] 
where $F:[0,T]\times C([0,T],\mathbb R^d)\times B \to \mathbb R$ and the asset follows the the SDE~\eqref{eq sde}. We denote by $F^{MC} : [0,T] \times \mathbb R^N \times B \to \mathbb R$ as the approximation of the price of the option calculated on a discretised path using $10^6$ Monte Carlo samples. 
\[
\mathcal E_{\text{integral}}^{\beta} =  \mathbb{E}^{N_{\text{trn}}} \left[  \sum_{t_k \in \pi^c}(t_{k+1}-t_{k}) \cdot \left| F^{MC}\left(t_k,(X_{t}^{\beta,\pi^f})_{t\in\pi^c};\beta\right) - \mathcal R[F]_{\theta}\left(t_k,(\mathcal X_{t}^{\beta,\pi^f})_{t\in\pi^c};\beta\right)\right|\right].
\]

\subsubsection{Integral error of the hedging strategy parametrisation}
We additionally evaluate the parametrisation of the hedging strategy in Algorithm~\ref{alg martingale representation} by calculating its absolute integral error
\[
\mathcal E_{\text{hedging}}^{\beta} =  \mathbb{E}^{N_{\text{trn}}} \left[ \sum_{t_k\in\pi^c} (t_{k+1}-t_{k}) \cdot \left| \nabla_{\omega} F^{MC}\left(t_k,(X_{t}^{\beta,\pi^f})_{t\in\pi^c};\beta\right) - \mathcal R[\nabla_{\omega} F]_{\phi}\left(t_k,(\mathcal X_{t}^{\beta,\pi^f})_{t\in\pi^c};\beta\right)\right|\right]
\]
where we denote by $\nabla_{\omega} F^{MC}\left(t_k,(X_{t}^{\beta,\pi^f})_{t\in\pi^c};\beta\right)$ as the approximation of the path derivative on a discretised path. It is approximated using $10^6$ Monte Carlo samples using~\eqref{eq vertical derivative}.

\subsubsection{Stochastic integral of the hedging strategy as a control variate} In Section~\ref{sec unbiased}, the discounted price is approximated by the estimator with low variance
\begin{equation*}
\label{eq for cv sec 2}		
M_t^{\text{cv},\phi}:= M_T - \int_t^T e^{-rs'}\big(\mathcal R_\phi[\nabla_\omega F]  \sigma\big)(s',(s', (X^\beta_{s'\wedge s})_{s\in[0,T]};\beta)\,dW_{s'}\,.
\end{equation*}
The correlation between the stochastic integral and $M_T$ should be close to $1$ in order to yield a good control variate (see ~\cite[Chapter 4.1]{glasserman2013monte}) and hence a good approximation of the hedging strategy. We provide the correlation between $M_T$ and the stochastic integral with $t=0$, and $x_0 = 1$.
\[
\rho\left(M_T, \int_0^T e^{-rs'}\big(\mathcal R_\phi[\nabla_\omega F]  \sigma\big)(s',(s', (X^\beta_{s'\wedge s})_{s\in[0,T]};\beta)\,dW_{s'}\right).
\]

\section{Numerical experiments}\label{sec results}

\subsection{Black Scholes model and lookback option}

Take a $d$-dimensional Wiener process $W$. 
We assume that we are given a symmetric, positive-definite matrix 
(covariance matrix) $\Sigma$ and a lower triangular matrix $C$ s.t.
$\Sigma=CC^*$.\footnote{For such $\Sigma$ we can always use Cholesky decomposition to find $C$.} 
The risky assets will have volatilities given by $\sigma^i$. 
We will (abusing notation) write $\sigma^{ij} := \sigma^i C^{ij}$, when we don't
need to separate the volatility of a single asset from correlations.
The risky assets under the risk-neutral measure are then given by 
\begin{equation} \label{eq BS SDE}
dS^i_t = rS^i_t \, dt + \sigma^i S^i_t \sum_j C^{ij} dW^j_t\,.
\end{equation}
Consider the lookback path-dependent payoff given by:
\[
g\left((S_t)_{t\in[0,T]}\right) = \left[\max_{t\in[0,T]}\sum_i S^i_t - \sum_i S^i_T \right]_+ 
\]
We take $d=2$, $r=5\%$, $\Sigma^{ii} = 1$, $\Sigma^{ij} = 0$ for $i\neq j$.
In our first two experiments where we learn the solution of the PPDE using Algorithms~\ref{alg orthogonal projection} and~\ref{alg martingale representation}, we will try two different volatility values, $\sigma^i = 30\%$ or $\sigma^i = 100\%$. That is, in these two experiments we solve separetely two PPDEs, rather than a whole family of PPDEs. In the third experiment we solve a parametric family of PPDEs by sampling $\sigma^i$ from $[0,0.4]$.

We take the maturity time $T=0.5$. We divide the time interval $[0,T]$
in 1000 equal time steps in the fine time discretisation, i.e. $\pi^f = \{0=t_0<\ldots<t_N=T, N=1\,000\}$,
and the following coarser time discretisation with 10 timesteps at which the network learns the prices:
$\pi^c = \{0=t_0<\ldots<t_{10}=T\} \subset \pi^f$.

We train our models with batches of 200 random paths $(s_{t_k}^i)_{n=0}^{N_\text{steps}}$ sampled from the SDE~\eqref{eq BS SDE} using Euler scheme
with $N_\text{steps}=100$. The assets' initial values $s_{t_0}^i$ are sampled from a lognormal distribution $S_0 \sim \exp ((\mu-0.5\sigma^2)\tau + \sigma\sqrt{\tau}\xi)$,
where $\xi\sim \mathcal{N}(0,1), \mu=0.08, \tau=0.1$.

In our experiments, Feed-forward networks consist of four hidden layers with 100 neurons each each of them 
followed by the activation function \texttt{ReLU}$(x) = \max(0,x)$. If we use LSTM networks, then we append a feed-forward network to each output of the LSTM with two layers and \texttt{ReLU} activation functions, to ensure that each output of the model can take values in $\mathbb R^m$.

All the path signatures are calculated up to the fourth iterated integral, and are calculated on the path or on the lead-lag transform of the path (see Appendix~\ref{sec path signatures}).

\subsubsection{Experiment 1 - Learning conditional expectation (Algorithm~\ref{alg orthogonal projection})}
We present results using the learning task~\eqref{eq learning task orthogonal projection},
and different combinations of network architectures and network
input. Results are shown in Table~\ref{table results conditional expectation}.

\begin{figure}[H]
\centering
  \includegraphics[width=0.49\linewidth]{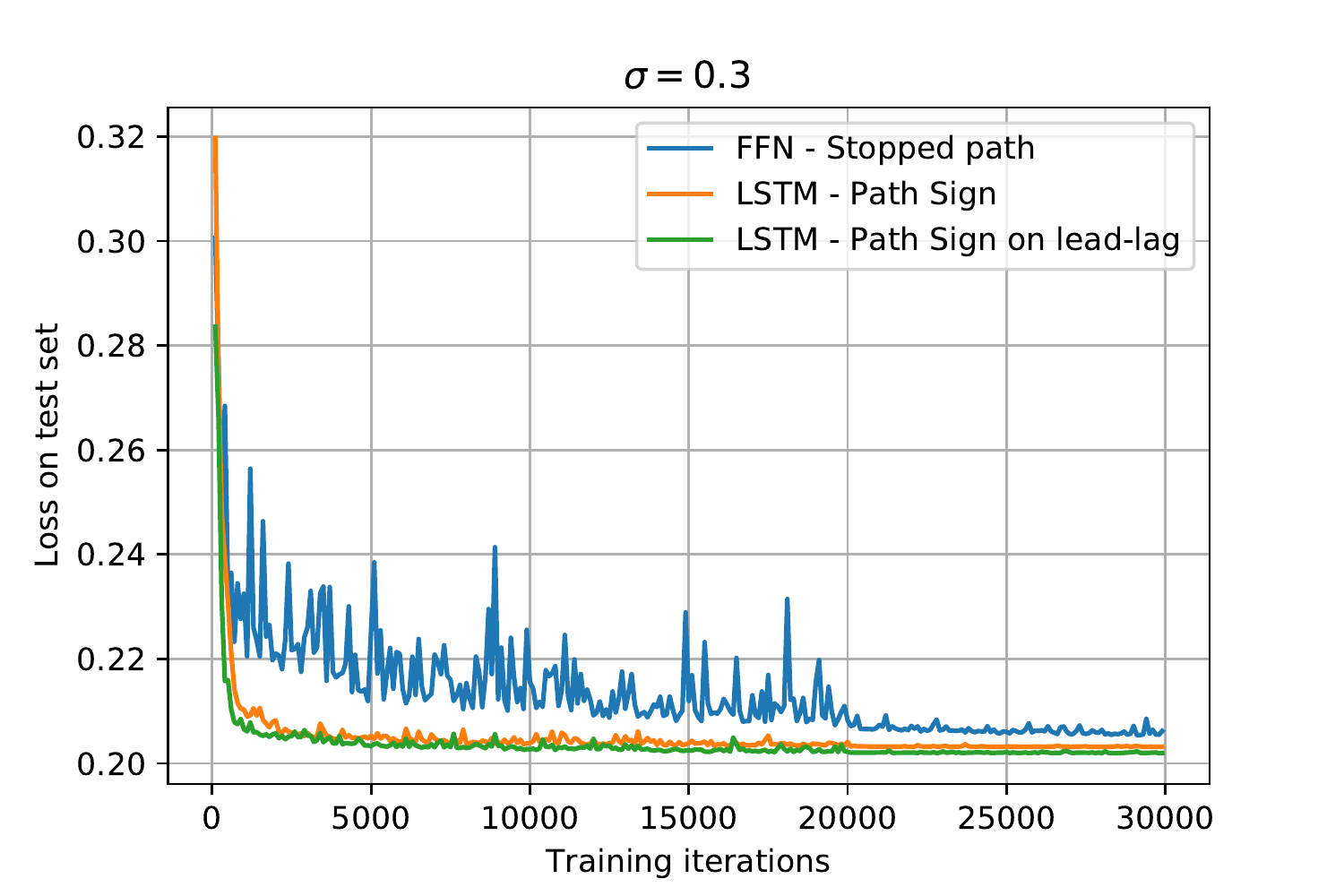}
  \includegraphics[width=0.49\linewidth]{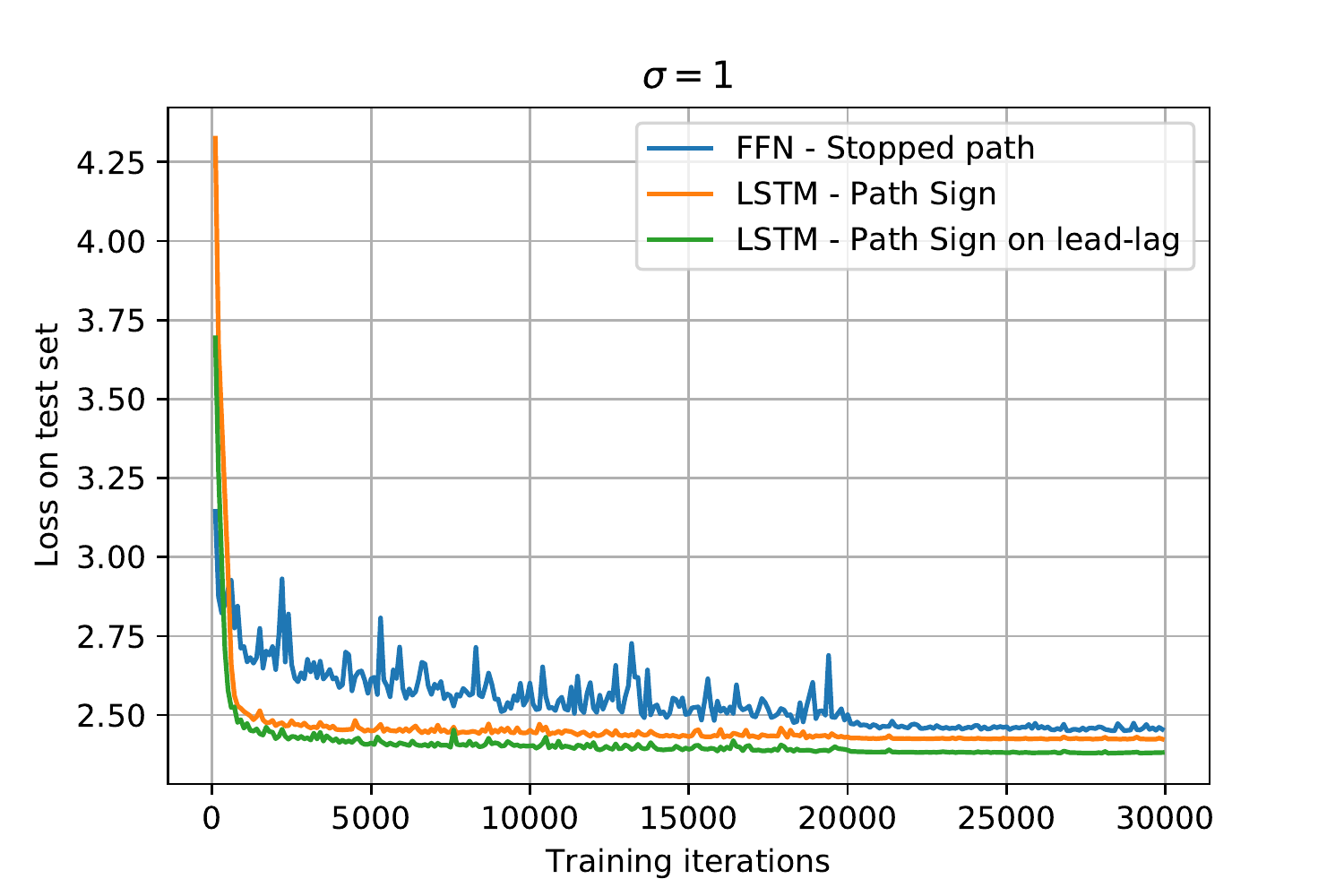}
\caption{Loss on test set in terms of training iterations}
\label{fig loss conditional expectation}
\end{figure}

LSTM networks combined with path signatures consistently give 
better results than feedforward networks. It is remarkable from 
Figure~\ref{fig loss conditional expectation} that one can see 
how path signatures start making a difference when the volatility 
of the underlying in the SDE~\eqref{eq BS SDE} increases. Observe that the loss depicted in Figure~\ref{fig loss conditional expectation} does not go down to 0. This comes from the fact that when we learn the conditional expectation, we are looking to find the $L^2$-distance between a random variable which is $\mathcal F_T$-measurable and its orthogonal projection on $\mathcal F_t\subsetneq \mathcal F_T$.

\begin{table}[H]
\caption{Evaluation of PPDE solver using algorithm~\ref{alg orthogonal projection}.}
\begin{tabular}{l|l|cc}
\hline
  Net. type & Net. input & $\sigma$ & $\mathcal E_{\text{integral}}$ \\
  \hline 
  FFN (baseline) & Path & $0.3$ & $8.64\times 10^{-3}$ \\ 
  LSTM & Path & $0.3$ & $5.88 \times 10^{-3}$  \\ 
  LSTM & Path sign & $0.3$ & $5.75 \times 10^{-3}$ \\ 
  \textbf{LSTM} & \textbf{Path sign on lead-lag transf} & $0.3$ & $\mathbf{4.45\times 10^{-3}}$ \\ \hline  
  FFN (baseline) & Path & 1 & $3.96\times 10^{-2}$ \\ 
   LSTM & Path & $1$ & $3.09\times 10^{-2}$   \\ 
   LSTM & Path sign & $1$ & $2.42 \times 10^{-2}$ \\ 
   \textbf{LSTM} & \textbf{Path sign on lead-lag transf} & $1$ & $\mathbf{1.73\times 10^{-2}}$ \\ 
\end{tabular}
\label{table results conditional expectation}	
\end{table}

\subsubsection{Experiment 2 - Learning Martingale Representation (Algorithm~\ref{alg martingale representation})}
We present results using the learning task~\eqref{eq loss function BSDE solver}.
Results are shown in Table~\ref{table Martingale representation}. We observe in Table~\ref{table results conditional expectation} that a combination of LSTM networks and path signatures provide the best results, and we restrict our experiments to this setting. We stress out that the obtained integral errors used in this method are slightly better than the errors we obtain after learning the conditional expectation (Table~\ref{table results conditional expectation}). We hypothesise that this is given due to higher structure of the solution contained in the loss function built to solve the BSDE.

\begin{table}[H]
\caption{Evaluation of PPDE solver using algorithm~\ref{alg martingale representation}.}
  \begin{tabular}{l|l|cccccc}
	\hline
   Net. type & Net. input & $\sigma$ & $\mathcal E_{\text{integral}}$ & $\mathcal E_{\text{hedging}}$ & $\rho$   \\
    \hline 
   LSTM & Path sign & $0.3$ & $4.5\times 10^{-3}$ & $2.02\times10^{-4}$ & $0.997$  \\ 
   \textbf{LSTM} & \textbf{Path sign lead-lag} & $\mathbf{0.3}$ & $\mathbf{3.86\times 10^{-3}}$ & $2.08\times10^{-4}$  & $\mathbf{0.998}$  \\ \hline 
   LSTM & Path sign & $1$ & $2.11\times 10^{-2}$ & $3.3\times 10^{-3}$  & $0.995$  \\ 
   \textbf{LSTM} & \textbf{Path sign lead-lag} & $\mathbf{1}$ & $\mathbf{1.60\times 10^{-2}}$ & $3.33\times 10^{-3}$ & $0.994$  \\ 
  \end{tabular}
  
  \label{table Martingale representation}
\end{table}

\subsubsection{Experiment 3 - Learning parametric PPDE}: We incorporate the volatility $\sigma^i, i=1,\ldots,d$
of the Black-Scholes model~\eqref{eq BS SDE} as an input to the networks in order to solve a parametric family of PPDEs using Algorithm~\ref{alg martingale representation}. $\sigma$ is uniformly sampled from $[0, 0.4]$. Results in Table~\ref{table parametric PPDE} show that parametric learning is shown to work as the error is consisten across the parameter range, with slightly higher errors for higher values of the volatility. 
 
\begin{table}[h]
\caption{Evaluation of parametric PPDE solver using using algorithm~\ref{alg martingale representation} and $\sigma$ as input to LSTM network}
\begin{tabular}{l|c|ccccc}
  \hline
  Method & $\sigma$ & Net. type & Net. input & $\mathcal E_{\text{integral}}$ & $\mathcal E_{\text{hedging}}$   \\
  \hline
  Martingale repr. & 0.05 & LSTM & Path sign lead-lag & $6.47\times 10^{-3}$ & $3.6\times 10^{-2}$ \\
  Martingale repr. & 0.1 & LSTM & Path sign lead-lag & $7.7\times 10^{-3}$ & $1.4\times 10^{-2}$  \\
  Martingale repr. & 0.15 & LSTM & Path sign lead-lag & $8.16 \times 10^{-3}$   & $8.1\times 10^{-3}$ \\  
  Martingale repr. & 0.2 & LSTM & Path sign lead-lag & $8.73\times 10^{-3}$  & $6.4\times 10^{-3}$ \\
  Martingale repr. & 0.25 & LSTM & Path sign lead-lag & $9.3\times 10^{-3}$   & $7.1\times 10^{-3}$ \\
  Martingale repr. & 0.3 & LSTM & Path sign lead-lag & $1.0\times 10^{-2}$ & $7.5\times 10^{-3}$ \\
  Martingale repr. & 0.35 & LSTM & Path sign lead-lag & $1.07\times 10^{-2}$ & $8.6\times 10^{-3}$ \\
 \end{tabular}
 \label{table parametric PPDE}
\end{table}

\subsection{Heston model and autocallable option}
In this experiment we consider the 1-dimensional Heston model with stochastic volatility
\[
\begin{split}
dS_t & = \sqrt{V_t}S_t dW_t^S, \quad S_0 = s_0 \\	
dV_t & = \kappa(\mu-V_t)dt + \eta \sqrt{V_t} dW_t^V, \quad V_0=v_0,\\
d \langle W^S, W^V \rangle_t & = \rho dt
\end{split}
\]
where we take $\kappa=3$, $\mu=0.3$, $\eta=1$, $\rho=0.6$, $v_0=s_0=1$.

We aim to price an autocallable option (see for example~\cite{alm2013monte}) on $(S_t)_{t\in[0,T]}$. Consider 
$m$ observation dates, $t_1<\ldots<t_m$, a barrier value $B$, premature payoffs $Q_1,\ldots,Q_m$, and a redempetion payoff $q(s)$.

Given a path $(S_t)_{t\in[0,T]}$ and the corresponding prices at the observation dates $S_{t_1}, \ldots, S_{t_m}$ then the discounted payoff of the  univariate autocallable option is given by:
\[
g ((S_t)_{t\in[0,T]}) =
\begin{cases}
	Q_j \quad \text{if} \quad S_{t_i} < B \leq S_{t_j} \quad \forall i<j, \\
	q(S_m) \quad \text{if} \quad S_{t_j} < B \quad \forall j	
\end{cases}
\]

The price of the autocallable option at current time $\tau$ is given by its expected 
(discounted) payoff. For example, if the option has 2 observation times, and $\tau=0$, then 
\[
\begin{split}
F(t, (s_{t\wedge\tau})_{\tau\in[0,T]}) & := \mathbb E[g\left((S_t)_{t\in[0,T]}\right) \vert (S_{t\wedge\tau})_{\tau\in[0,T]} = (s_{t\wedge\tau})_{\tau\in[0,T]}] \\
& =  Q_1\mathbb E[\mathds{1}_{B \leq S_{t_1}/S_0}] + Q_2 \mathbb E[\mathds{1}_{B > S_{t_1}/S_0}\mathds{1}_{B \leq S_{t_2}/S_0}] + \mathbb E[q(S_T) \mathds{1}_{B > S_{t_1}/S_0}\mathds{1}_{B > S_{t_2}/S_0}]\\
\end{split}
\]

We use the parameters in table~\ref{table autocall} for the option payoff.
\begin{table}[H]\label{table autocall}
\caption{Parameters of the autocallable option}	
\begin{tabular}{l|c}
\hline
  Parameter & Value \\
  \hline
  Maturity & $T = 0.5$ years \\
  Barrier & $B=1.02$ \\
  No. of observation dates & $m=2$ \\
  Observation dates & 2,4 months \\
  Premature payoffs & $Q_1 = 1.1, Q_2=1.2$ \\
  Redemption payoff & $q(s) = 0.9s$ \\
\end{tabular}
\end{table}

\begin{figure}
\centering
  \includegraphics[width=0.8\linewidth]{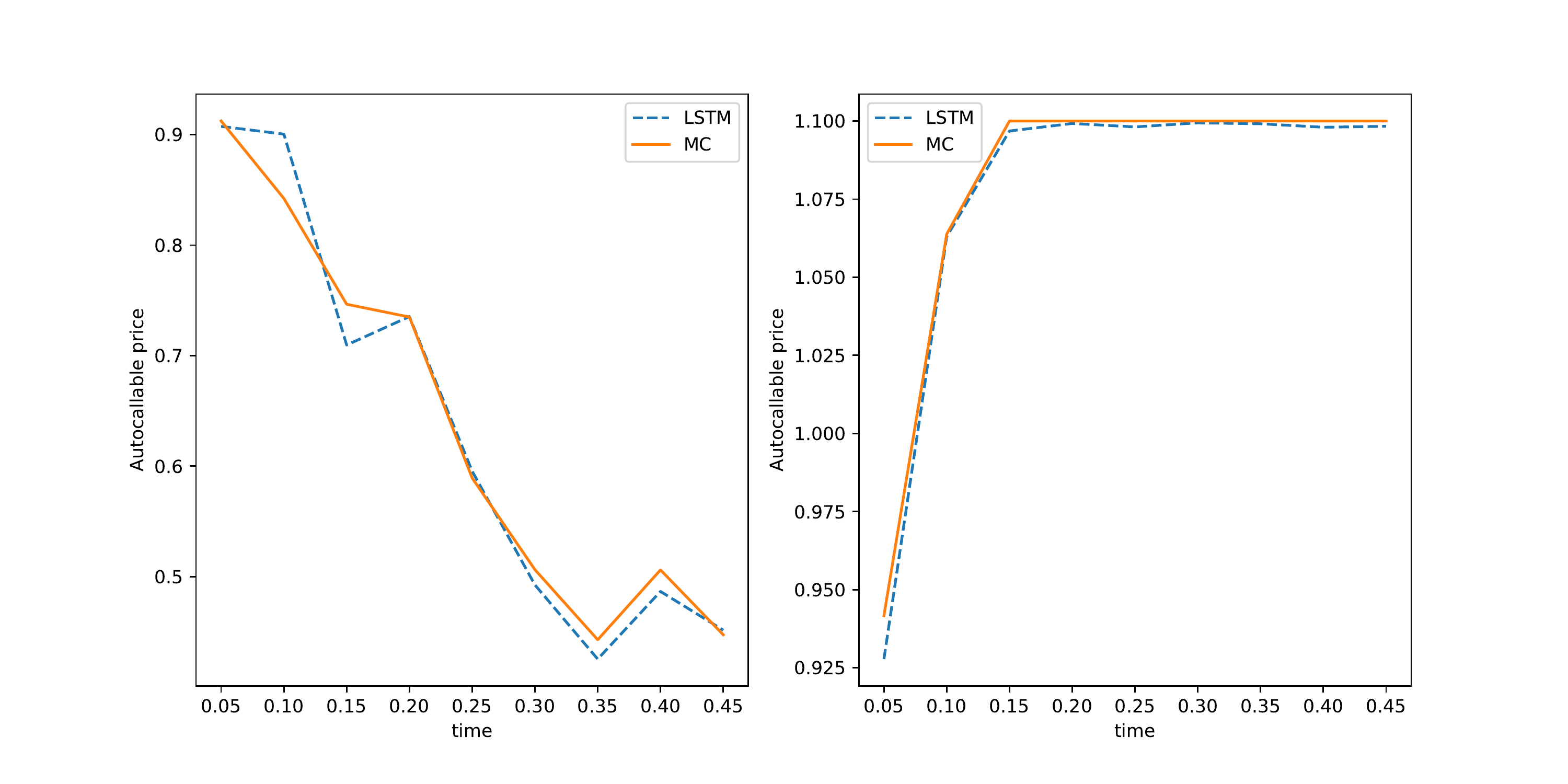}
\caption{Monte Carlo price and predicted price using LSTM networks trained using the Martingale representation of the price for two different paths of the underlying share}
\label{fig heston price}
\end{figure}

Results of the approximation of the PPDE using Algorithm~\ref{alg martingale representation} are provided in Table~\ref{table autocall results}. Figure~\ref{fig heston price} displays the Monte Carlo approximation of the PPDE solution and the LSTM approximation

\begin{table}[H]\label{table autocall results}
\caption{Evaluation of PPDE solver using Heston model, autocallable option and algorithm~\ref{alg martingale representation}.}	
\begin{tabular}{l|ccccc}
  \hline
  Method & Net. type & Net. input & $\mathcal E_{\text{integral}}$ & $\mathcal E_{\text{hedging}}$ & $\rho$  \\
  \hline
  Martingale repr. & LSTM & Path sign lead-lag &  $1.4 \times 10^{-2}$ & $1.8 \times 10^{-2}$ & $0.948$ 
 \end{tabular}
\end{table}

%
%

\section{Conclusions}
In this paper we implement two numerical methods to approximate the solution of a parametric family of PPDEs. 
\begin{equation*}
\begin{split}
& \bigg[\partial_t F + b \nabla_\omega F + \frac{1}{2}\text{tr}\left[\nabla_{\omega}^2 F \sigma^*\sigma\right] - rF\bigg](t,\omega;\beta)  = 0\,, \\
& F(T,\omega;\beta)  = g(\omega;\beta)\,,\,\,\, t \in [0,T]\,,\,\, \omega \in C([0,T];\mathbb R^d)\,,\,\,\beta \in B\,.
\end{split}
\end{equation*} 
by using the probabilistic representation of $F(t,\omega; \beta)$ given by Feynman-Kac formula. This representation allows to tackle the problem of solving the PPDE, as pricing an option where the underlying asset follows a certain SDE in the risk neutral measure. This is the setting of our numerical experiments, where we price lookback and autocallable options using properties of the conditional expectation (algorithm~\ref{alg orthogonal projection}) or the martingale representation of the price (algorithm~\ref{alg martingale representation}). In the latter algorithm, we parametrise $\nabla_{\omega}F$ by a neural network, which in some models provides the hedging strategy. 
We combine path signatures to encode the information in $\omega$ in some finite dimensional structure together with LSTM networks to model non-anticipative functionals, that in our experiments provide a higher accuracy than Feed Forward Networks.

\bibliographystyle{abbrv}
\bibliography{deep_pde.bib}

\newpage
\appendix
\section{Functional Ito Calculus}\label{sec Func Ito Calculus}
In this section we define the notion of path-dependent PDE.  and we review the 
theory that it relies on, Functional Ito calculus~\cite{dupire2019functional, bally2016stochastic}. 
Consider the space of c\` adl\` ag paths in $[0,T]$, $D([0,T], \mathbb R^d)$. 
The space of stopped paths is the quotient space 
\[
\Lambda_T := \left([0, T]\times D([0,T], \mathbb R^d)\right)/\sim
\]
defined by the equivalence
relationship
\[
(t, \omega) \sim (t', \omega') \Leftrightarrow t=t' \text{ and } 
(\omega_{s\wedge t})_{s\in[0,T]} = (\omega'_{s\wedge t})_{s\in[0,T]}.
\] 
Consider a functional $F: \Lambda_T \rightarrow \mathbb R$. The continuity of $F$ is
defined with respect to the metric in $\Lambda_T$:
\[
d_{\infty} ((t,\omega), (t', \omega')) = \sup_{s\in[0,T]}\vert \omega_{s\wedge t} - \omega'_{s\wedge t'} \vert + |t-t'| 
\]

Furthermore, a functional $F$ is \textit{boundedness preserving} if for every compact 
set $K \subset \mathbb R^n$, and $\forall t_0\in[0,T]$ there exists a constant $C>0$
depending on $K, t_0$ such that 
\[
\forall t\in[0,t_0], \quad \forall (t,\omega) \in \Lambda_T, \quad  (\omega_s)_{s\in[0,t]}\subset K
\Rightarrow F(t,\omega) < C.
\] 

And finally we define, when the limit,
exists the horizontal derivative of a functional $F$
\[
\partial_t F(t,\omega) := \lim_{h\rightarrow 0^+} \frac{F(t+h, \omega) - F(t,\omega)}{h} 
\]
and the vertical derivative
\begin{equation}\label{eq vertical derivative}
\nabla_{\omega} F(t,\omega) := (\partial_i F(t,\omega), i=1,\ldots,d) \in \mathbb R^d
\end{equation}
where
\[
\partial_i F(t,\omega) := \lim_{h\rightarrow 0} \frac{F(t,(\omega_{t\wedge s})_{s\in[0,T]} + he_i \mathbf{1}_{[t,T]}) - F(t,(\omega_{t\wedge s})_{s\in[0,T]})}{h}  
\]
with $e_i$ the canonical basis of $\mathbb R^d$. The functional $\nabla_{\omega} F: \Lambda_T \rightarrow \mathbb R^d$ is well defined in the quotioent 
space $\Lambda_T$, and therefore one can calculate higher order path derivatives 
by repeating the same operation when the 
limit exists. 

A functional $F: \Lambda_T \rightarrow \mathbb R$ belongs to $\mathbb C^{1,2}$ if:
\begin{enumerate}[i)]
\item It is continuous.
\item It is boundedness preserving.
\item It has continuous, boundedness preserving derivatives $\partial_t F, \nabla_{\omega} F, \nabla_{\omega}^2 F$.	
\end{enumerate}

\section{The signature of a path}
\label{sec path signatures}
Iterated integrals of piece-wise regular multi-dimensional paths 
were first studied by K.T. Chen~\cite{chen1957integration, chen1958integration},
and the study of their properties was extended for continuous paths of bounded variation in~\cite{hambly2010uniqueness}. 

Given a $d$-dimensional path $X: [a,b]\rightarrow \mathbb R^d$, we denote 
the coordinate paths $(X_t^1, \ldots,X_t^d)$ where each $X^i:[a,b]\rightarrow \mathbb R$.

For each $i=1,\ldots,d$ the \textit{first iterated integral} of the $i$-th coordinate path is
\[
S(X)^i_{a,t} := \int_{a<s<t} dX^i_s = X^i_t-X^i_a.
\]

Note that since $S(X)^i_{[a,\cdot]}: [a,b]\rightarrow \mathbb R$ is also a continuous path, we 
can integrate it again along any of the coordinate paths to obtain the 
\textit{second iterated integral}: for any $i,j$,
\[
S(X)^{i,j}_{a,t} := \int_{a<s<t} S(X)^i_{a,s} dX^j_s = \int_{a<r<s<t} dX^i_r dX^j_s.
\]
This process can be repeated to obtain any coordinate of the $k$-th iterated integral:
\[
S(X)^{i_1,\ldots,i_k}_{a,t} = \int_{a<t_1<t_2<\ldots<t_k<t} dX_{t_1}^{i_1}\ldots dX_{t_k}^{i_k}
\]
We introduce the concept of tensor algebera $T((\mathbb R^d))$, which is where the signature of an $\mathbb R^d$-valued path takes its values.  
\begin{definition}
	Consider the basis of a vector space $E$ given by $\{e_1,e_2,\ldots,e_d\}$, and the successive tensor powers $E^{\otimes n}$, which can be identified with the space of degree $n$ in $d$ variables
	\[
	\sum_{i_1\ldots i_n\in\{1,\ldots,d\}} \lambda_{i_1,\ldots,i_n}e_{i_1}\ldots e_{i_n}
	\] 
	The tensor algebra space denoted by $T((E))$ is then defined as
	\[
	T((E)) := \left\{(a_0,a_1,\ldots,a_n,\ldots)\, | \, \forall n\geq0, a_n \in E^{\otimes n} \right\}
	\]
	The $n$-th truncated tensor algebra space is
	\[
	T^n(E) := \bigoplus_{i=1}^n E^{\otimes n}
	\]
\end{definition}
Thus, the $k$-th iterated integral of $X$ can be defined as
\[
S(X)_{a,t}^{(k)} = \int_{a<t_1<t_2<\ldots<t_k<t} dX_{t_1} \otimes \ldots \otimes dX_{t_l} \in T^k(\mathbb R^d)
\]
\begin{definition}\label{def signature}
Let $\mathcal I$ denote all the set of multi-indices $(i_1,\ldots,i_k)$ with $k\geq 0$
and $i_j \in \{1,\ldots,d\}$. 
The signature of $X:[a,b]\rightarrow \mathbb R^d$ is an element of the tensor algebra $T((\mathbb R^d))$,
\[
\text{Sig}_{a,b}(X) = (S(X)^{I}_{a,b})_{I\in\mathcal I} = (1,S(X)_{a,t}^{(1)}, S(X)_{a,t}^{(2)},\ldots)\in T((\mathbb R^d)).
\]
\end{definition}

\subsection{The signature of a data stream}
So far we have built the path signature on continuous trajectories. In financial data,
one normally deals with data streams, i.e. trajectories defined by a sequence
of time points $(x^{\pi}_{t_i})_{i=1,\ldots,N}$. The common approach to define 
the signature of this data stream is via the iterated integrals of its piece-wise
linear interpolation.

\subsection{Machine Learning and the signature method}
For a premier on the use of the signature in Machine Learning, we refer the reader
 to~\cite{chevyrev2016primer} and for a rigorous treatment of the signature
properties the reader can refer to~\cite{hambly2010uniqueness}. We state however the following 
two properties that motivate using the signature of a path in Machine Learning.

\begin{enumerate}[i)]
\item The terms of the signature decay in size factorially~\cite[Lemma 2.1.1]{lyons2007differential}, i.e.
\[
\left\Vert S(X)_{a,t}^{(k)}\right\Vert \leq \frac{C(X)^k}{k!}
\]
where $C(X)$ depends on $X:[a,b]\rightarrow \mathbb R^d$ and $\Vert \cdot \Vert$ is a tensor norm
in $T^k(\mathbb R^d)$. As a consequence of this, it is usual in machine learning to 
truncate the signature up until a certain depth $n$, obtaining
\[
\text{Sig}^{(n)}_{[a, b]}(X) = (1,S(X)_{a,t}^{(1)}, S(X)_{a,t}^{(2)},\ldots, S(X)_{a,t}^{(n)})
\]

\item The signature is rich enough that every continuous function of the path can 
be approximated by a linear function of its truncated signature. More precisely, 
the universality result given in Theorem 3.1 of ~\cite{levin2013learning} tells us that any continuous functional on the paths can be approximated up until any accuracy $\varepsilon$ by a linear combination of the coordinates of the truncated path signature $\text{Sig}^{(n)}_{[a, b]}(X)$, for some $n:=n_{\varepsilon}$. 

\end{enumerate}

\subsection{Lead-lag transform}
We finally introduce the lead-lag transform of a a data stream ~\cite{Flint_2016}, that  to write Ito integrals 
as linear functionals on the signature of the lead-lag transformed path.

More specifically, given a stream of data $(x^{\pi}_{t_i})_{i=1,\ldots,N}$, then 
we define the lead-transformed stream as
\[
x^{\pi, \text{lead}}_j = 
\begin{cases}
	x^{\pi}_{t_i} \text{ if } j=2i \\
	x^{\pi}_{t_i} \text{ if } j=2i-1 
\end{cases}
\] 
and the lag-transformed stream as
\[
x^{\pi, \text{lag}}_j = 
\begin{cases}
	x^{\pi}_{t_i} \text{ if } j=2i \\
	x^{\pi}_{t_i} \text{ if } j=2i+1 
\end{cases}.
\] 
The resulting lead-lag transformed stream is:
\[
(x^{\pi, \text{lead-lag}}_{t_i})_{i=1,\ldots,2N} = (x^{\pi, \text{lead}}_{t_i}, x^{\pi, \text{lag}}_{t_i})_{i=1,\ldots,2N}.
\]

\section{Deep Neural Networks for function approximation}
\label{sec network architecture}
\subsection{Feedforward neural networks}
\label{sec ffn}

A fully connected artificial neural network is given by 
by a composition of affine transformations and non-linear activation functions. 
Fix $L$ as the number of layers, then the space of parameters of the network is 
given by
\[
\Pi = (\mathbb R^{l^1\times l^0}\times \mathbb R^{l^1}) \times (\mathbb R^{l^2\times l^1}\times \mathbb R^{l^2})\times\cdots\times (\mathbb R^{l^L\times l^{L-1}}\times \mathbb R^{l^L})\,,
\]	
hence if we denote the parameters of a network by 
\[
\theta := ((W^1,b^1), \ldots, (W^L, b^L)) \in \Pi\,.
\]
and by denoting the $i$-th network layer by $M^i$ such that
\[
M^i(z_{i-1}) = \varphi^i(W^i z_{i-1} + b^i) 
\]
with $\varphi$ being a non-linear activation function such as $\tanh$ or the sigmoid, 
then the reconstruction of $\mathcal R_{\theta} : \mathbb R^{l^0} \to \mathbb R^{l^L}$ 
can be written recursively by 

\begin{equation}
\label{eq FFN}
y: = \mathcal R_{\theta}(z_0) = W^L z^{L-1} + b^L\,, \,\,\,\, 
z^k = M^k(z_{k-1})\,.
\end{equation}

\subsection{Long Short Term Memory Networks}
Long Short Term Memory (LSTM) networks \cite{hochreiter1997long} are an example of Recurrent Neural Networks, which are useful when the input is a sequence of points
\[
\{x_0, x_1, \ldots, x_n\}.
\]
 
Each element $x_t$ of the input sequence is fed to the Recurrent Neural Network which
in addition to returning an output $y_t$, also stores some information (or
hidden state) $a_t$ that is used to perform computations in the next step:
More formally, 
\[
\mathcal R_{\theta} (x_{t}, a_{t-1}) = (y_{t}, a_{t}).
\]

LSTM networks are designed to tackle the problem of exploding or vanishing gradients that 
plain RNN suffer from (see\cite{bengio1994learning}). They do this by regulating 
the information carried forward by the hidden state given each input of the sequence,
using the so-called \textit{gates}. Specifically, the operations performed 
for the $i$-th element of the sequence $x_i$, receiving the hidden state $a_{t-1}:= (h_{t-1},c_{t-1})$ are:
\begin{align*}
i_t &= \sigma(W_{xi}x_t + b_{xi} + W_{hi}h_{t-1} + b_{hi}) \\
f_t &= \sigma(W_{xf}x_t + b_{xf} + W_{hf}h_{t-1} + b_{hf}) \\
g_t &= \tanh(W_{xg}x_t + b_{xg} + W_{hg}h_{t-1} + b_{hg}) \\
o_t &= \sigma(W_{xo}x_t + b_{xo} + W_{ho}h_{t-1} + b_{ho}) \\
c_t &= f_t \odot c_{t-1} + i_t\odot g_t \\
h_t & = o_t \odot \tanh(c_t)				
\end{align*}
in addition, since $h_t\in (0,1)$, we add a linear layer 
\[
y_t = W_{hy}h_t + b_{hy}.
\]
Where $x_t\in \mathbb R^d, W_{x*} \in \mathbb R^{k\times d}, b_{x*}\in \mathbb R^k$, $k\in \mathbb Z_+$, and $\odot$ is the element-wise multiplication of two vectors.

\end{document}